%% file: main.tex
\documentclass[acmsmall,screen]{acmart}

\input{setup}
\title{GUIMigrator: Semantics-Preserving Transpilation from Android XML to Compose and SwiftUI}

\setcopyright{cc}
\setcctype{by}
\acmDOI{10.1145/3808215}
\acmYear{2026}
\acmJournal{PACMSE}
\acmVolume{3}
\acmNumber{FSE}
\acmArticle{FSE208}
\acmMonth{7}
\acmSubmissionID{fse26mainb-p3073-p}
\received{2026-02-23}
\received[accepted]{2026-03-24}

\begin{document}

\author{Yi Gao}
\orcid{0009-0000-2554-2381}
\affiliation{%
  \institution{Zhejiang University}
  \city{Hangzhou}
  \country{China}
}
\affiliation{%
  \institution{the Hangzhou High-Tech Zone (Binjiang) Institute of Blockchain and Data Security}
  \city{Hangzhou}
  \country{China}
}
\email{gaoyi01@zju.edu.cn}

\author{Xing Hu}
\orcid{0000-0003-0093-3292}
\affiliation{%
  \institution{Zhejiang University}
  \city{Hangzhou}
  \country{China}
}
\email{xinghu@zju.edu.cn}

\author{Xiaohu Yang}
\orcid{0000-0003-4111-4189}
\affiliation{%
  \institution{Zhejiang University}
  \city{Hangzhou}
  \country{China}
}
\email{yangxh@zju.edu.cn}

\author{Xin Xia}
\orcid{0000-0002-6302-3256}
\authornote{Corresponding Author}
\affiliation{%
  \institution{Zhejiang University}
  \city{Hangzhou}
  \country{China}
}
\affiliation{%
  \institution{the Hangzhou High-Tech Zone (Binjiang) Institute of Blockchain and Data Security}
  \city{Hangzhou}
  \country{China}
}
\email{xin.xia@acm.org}

\input{abstract}

\maketitle

\input{introduction}
\input{motivation}
\input{approach}
\input{evaluation}

\input{discussion}
\input{related_work}
\input{conclusion}

\begin{acks}
This work was supported by the National Key R\&D Program of China (No. 2024YFB4506400).
\end{acks}

\bibliographystyle{ACM-Reference-Format}
\bibliography{main}
\end{document}

%% file: setup.tex
\usepackage[ruled,linesnumbered]{algorithm2e}
\usepackage{multirow}
\usepackage{listings}
\usepackage{booktabs}
\usepackage{multicol}
\usepackage{tabularx}
\usepackage{adjustbox}
\usepackage{subfig}
\usepackage{subcaption}
\usepackage{makecell}
\usepackage{pifont}
\usepackage{enumitem}
\usepackage{calligra}
\usepackage{mdframed}
\usepackage{fontawesome}
\usepackage{amsmath}
\usepackage{amsfonts}
\usepackage{graphicx}
\usepackage{textcomp}
\usepackage{xcolor}
\usepackage{balance}
\usepackage{caption}
\usepackage{wrapfig}
\usepackage{colortbl}
\usepackage{forest}
\usepackage{array}
\usepackage[most]{tcolorbox}

\usepackage{wrapfig}

\usepackage{booktabs}
\usepackage{colortbl}
\usepackage{xcolor}

\usepackage{forest}
\usepackage[most]{tcolorbox}

\def\BibTeX{{\rm B\kern-.05em{\sc i\kern-.025em b}\kern-.08em
    T\kern-.1667em\lower.7ex\hbox{E}\kern-.125emX}}

\lstdefinestyle{codeStyle}{
    basicstyle=\small\ttfamily,
    commentstyle=\color{green},
    keywordstyle=\color{blue},
    stringstyle=\color{blue},
    showstringspaces=false,
    breaklines=true,
    frame=lines,
    backgroundcolor=\color{white},
    captionpos=b,
    aboveskip=10pt,
    belowskip=10pt
}
\definecolor{lightgreen}{rgb}{0.9,1,0.9}
\definecolor{grayheader}{gray}{0.85}
\definecolor{codebackground}{RGB}{240,240,240} %

\newcommand{\appname}{{\sc GUIMigrator}\xspace}
\newcommand{\apps}{\textit{apps}\xspace}

%% file: abstract.tex
\begin{abstract}
Constructing user interfaces (UIs) is one of the most resource-intensive tasks in mobile development, often consuming more than half of overall effort. 
Although declarative frameworks such as Jetpack Compose (Android) and SwiftUI (iOS) have become mainstream, the majority of existing Android apps still rely on legacy XML-based layouts. 
Migrating these UIs to declarative paradigms is essential for maintainability and cross-platform reuse, but manual migration is costly, error-prone, and difficult to scale.
We present \appname, a semantics-preserving framework that automates the migration of Android XML-based UIs to Jetpack Compose and SwiftUI. 
We design the \emph{Semantic UI Transpiler (SUT)}, which abstracts layout structures and resource semantics from legacy XML and systematically re-expresses them using the component abstractions and idioms of modern declarative frameworks. 
This design ensures that migrated UIs preserve both visual fidelity and functional equivalence, while generating idiomatic, compilable code that maintains cross-platform consistency with minimal manual intervention.
By separating semantic interpretation from platform-specific realization, \appname provides a deterministic yet extensible basis for cross-platform modernization, avoiding the unpredictability of purely generative approaches. 
We evaluate \appname on 31 open-source applications across ten domains.
Results show that \appname achieves high migration completeness and strong visual similarity (81.9\% SSIM on Jetpack Compose and 78.2\% on SwiftUI on average), while maintaining substantially higher project-wide semantic coherence (PSC) than modern LLM baselines.
In addition, \appname reduces manual development effort by over 90\%.
These findings demonstrate that \appname provides an effective and practical solution for reusing Android UIs across modern declarative frameworks, advancing automated cross-platform UI development.

\end{abstract}

\ccsdesc[500]{Software and its engineering~Software evolution}
\ccsdesc[500]{Software and its engineering~Reusability}

\keywords{UI Transpiler, UI Migration, SwiftUI, Jetpack Compose, Android, iOS}

%% file: introduction.tex
\section{Introduction}
\label{sec:introduction}
Mobile applications (\apps) have become an indispensable part of modern life, supporting communication, commerce, entertainment, and productivity~\cite{chen2018ui,beltramelli2018pix2code}. 
To improve user experiences, companies usually provide \apps with attractive and consistent UIs on different platforms~\cite{biorn2020empirical,el2017taxonomy,el2016enhanced,vidmark2025performance,punguil2025analisis,mahmoud2024trans,abbas2024enhancing}, such as Android~\cite{android} and iOS~\cite{ios}.
However, UI development is among the most resource-intensive tasks, often consuming over half of the total engineering effort~\cite{feng2022auto}.
Recently, declarative frameworks such as Jetpack Compose (Android)\cite{compose} and SwiftUI (iOS)\cite{swiftui} have been promoted as the standard for maintainable and expressive UIs.
These frameworks represent a paradigm shift from legacy XML layouts, which still dominate many Android \apps.
As mobile ecosystems evolve, migrating legacy XML UIs to modern declarative paradigms is essential for maintainability and reuse. 
However, manual migration is costly, error-prone, and impractical at scale—especially for organizations maintaining large, long-lived Android codebases~\cite{chen2026every}.

Several research and industrial efforts have attempted to reduce the burden of UI development~\cite{beltramelli2018pix2code,talebipour2021ui,zhang2024llamatouch,ahmed2025inferring,jiang2025screencoder,rodriguez2025starvector,lu2025misty}. 
Computer-vision–based approaches translate UI screenshots into code, but they are fragile to minor variations: even small changes in resolution, style, or occlusion often lead to misrecognized elements and unusable layouts.
Cross-platform frameworks such as Flutter~\cite{flutter} and React Native~\cite{react} mitigate duplication by enabling code reuse, but they require developers to redesign UIs in an entirely different paradigm, sacrificing native fidelity and breaking compatibility with existing codebases. 
More recently, large language models (LLMs) have been explored for code translation, showing promise in generating declarative UI code. 
Nevertheless, LLMs frequently hallucinate non-existent components, misorder attributes, or generate unstable code that fails to compile—yielding results that are neither predictable nor reproducible. 

Migrating legacy Android XML layouts into modern declarative frameworks is a necessary but non-trivial task.
The task is complicated by several semantic gaps between paradigms:
First, \textit{Language heterogeneity}: Android XML encodes UIs in an imperative, tree-structured format, while declarative frameworks (Compose, SwiftUI) rely on composable functions and nested containers. Direct syntactic translation fails to preserve layout intent.
Second, \textit{Resource incompatibility}: Android resources such as colors, fonts, and vector graphics do not directly map to iOS or Compose counterparts, requiring systematic adaptation rather than simple substitution.
Finally, \textit{Complex layouts}: Advanced constructs like \texttt{ConstraintLayout} encode relational constraints among views that lack direct equivalents in SwiftUI, demanding careful semantic approximation to preserve visual fidelity.
These challenges reveal that UI migration cannot be solved by syntactic rewriting alone. 
Instead, it requires a semantics-preserving transformation that bridges structural, visual, and resource-level differences while producing idiomatic code for the target platform.

\noindent \textbf{Our Insight.} Despite the syntactic and semantic differences, declarative frameworks such as Jetpack Compose and SwiftUI provide comparable abstractions for layout, views, and resources. 
This observation suggests that existing Android XML-based UIs can be systematically reused by migrating them into equivalent declarative constructs, provided that the migration process captures layout semantics and adapts resources appropriately. 

In this paper, we present \appname, a semantics-preserving approach for migrating legacy Android XML-based UIs to modern declarative frameworks.
\appname formulates UI migration as a semantic transpilation problem across heterogeneous UI paradigms.
\appname represents a semantic-to-declarative migration methodology that bridges imperative XML-based layout semantics and declarative UI frameworks through a reusable semantic layer, enabling cross-paradigm migration without manual one-to-one widget mappings.
At its core is the \emph{Semantic UI Transpiler (SUT)}, which constructs a platform-agnostic intermediate representation that encodes both structural hierarchies and semantic relationships extracted from Android XML.
\appname performs migration through a structured three-step pipeline:
\ding{182} \textbf{Resource Parsing and Adaptation}, which resolves and normalizes Android-specific assets;
\ding{183} \textbf{UI Transpilation}, which interprets the semantic roles of layouts and views and maps them into equivalent declarative constructs; and
\ding{184} \textbf{Target UI Generation}, which instantiates the transpiled models into platform-specific code templates, producing compilable SwiftUI and Jetpack Compose code.

By following this structured pipeline, \appname enables efficient and accurate modernization of legacy Android UIs while reducing the manual burden of redeveloping UIs across platforms.
This new paradigm separates UI semantic modeling from platform-specific realization, providing a deterministic and extensible way to preserve intent across heterogeneous frameworks while avoiding the unpredictability of LLM-based generation and the redesign cost of cross-platform frameworks.

We evaluate \appname on 31 open-source Android applications spanning ten domains. 
Our study measures migration quality in terms of correctly transpiled layouts and views, visual fidelity between pre- and post-migration screenshots, and code correctness. 
Across 1,027 XML layouts, \appname achieves substantially higher similarity than GPT-5.2 and DeepSeek-3.2 baselines, while maintaining stable and compilable code. 
A controlled user study further confirms that \appname reduces developer effort by over an order of magnitude in both code changes and development time compared with manual implementation. 
Finally, performance measurements show that the entire dataset can be migrated within a few seconds, demonstrating the efficiency and practicality of our approach.
This work makes the following contributions: 

\begin{itemize}[leftmargin=*]
    \item We propose the first semantics-preserving UI migration method that formulates legacy Android XML layouts as a semantic transpilation problem, enabling reusable migration across modern declarative frameworks (Jetpack Compose and SwiftUI). 
    \item We introduce a semantic transpilation methodology based on the Semantic UI Transpiler (SUT), which captures cross-paradigm UI semantic invariants and models platform-agnostic layout semantics, enabling systematic translation into platform-specific declarative UI code.
    \item We evaluate \appname on 31 open-source applications from ten domains. Results show that \appname achieves high migration completeness and visual fidelity, outperforms GPT-5.2 and DeepSeek-V3.2 baselines, and reduces manual development effort by over an order of magnitude. 
\end{itemize}

%% file: motivation.tex
\section{Motivation Example}
\label{sec:motivation}

\begin{figure}
\centering
\includegraphics[width=0.9\linewidth]{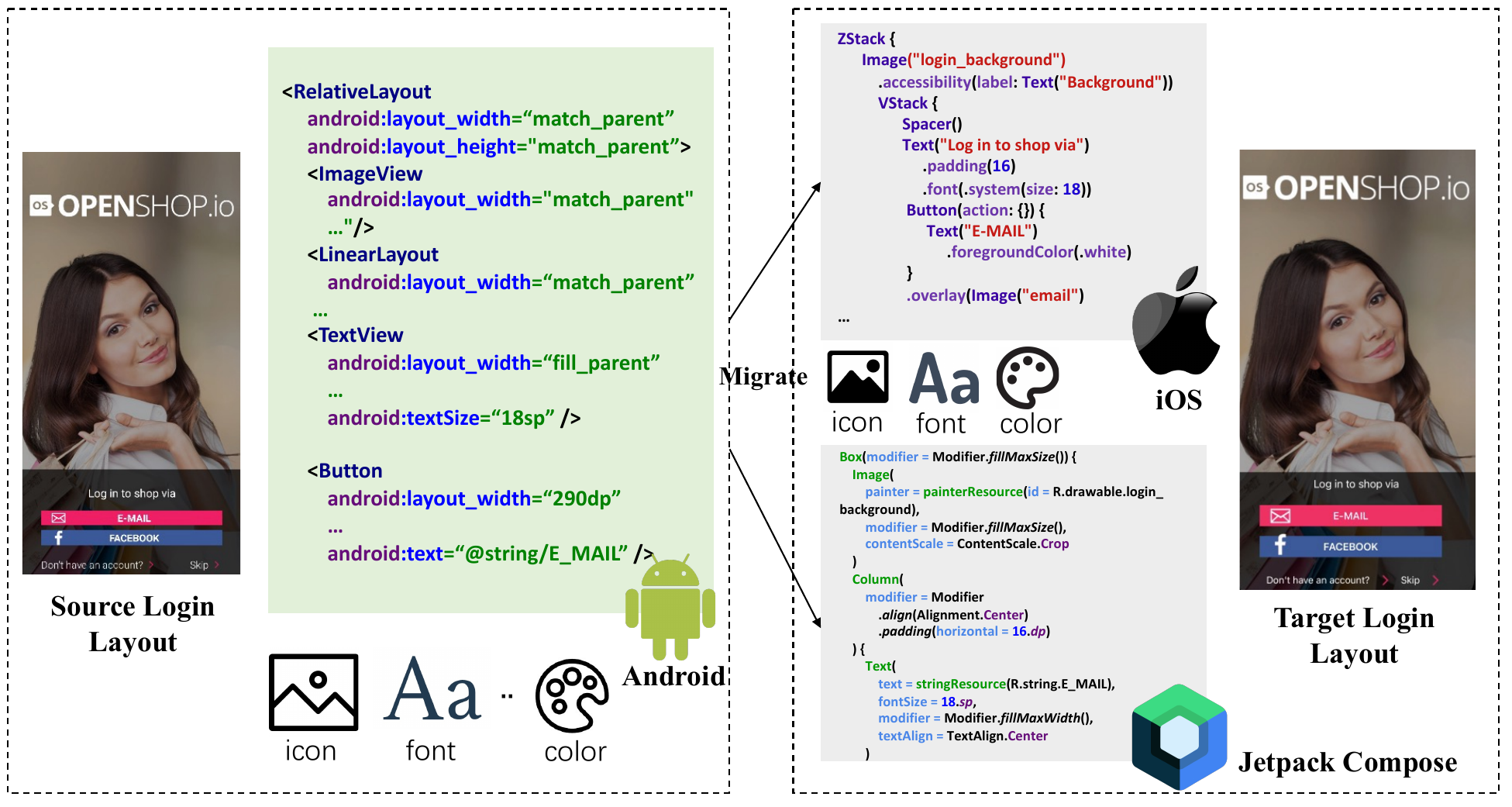}
\caption{Example of cross-platform UI migration. 
A legacy Android XML layout is converted into SwiftUI (iOS) and Jetpack Compose (Android), illustrating the structural gap between imperative XML and declarative frameworks and motivating a semantics-preserving transpilation approach.}
\label{fig:motivation}
\end{figure}

\noindent\textbf{Use Case.} 
Consider a mobile development company that wishes to modernize its legacy Android applications.
Most of its UIs are defined in XML, but the company now wants to (i) adopt Jetpack Compose for new Android versions and (ii) support iOS with SwiftUI to reach more users.
The straightforward option is to rebuild UIs manually, requiring developers proficient in both platforms.
Alternatively, cross-platform frameworks (e.g., Flutter, React Native) provide portability but sacrifice native fidelity and introduce performance overhead.
Recent approaches leveraging screenshots or LLMs can partially reuse existing designs, but they often fail to respect semantic UI constraints, resulting in inaccurate layouts or hallucinated code.
These limitations make the direct reuse of legacy Android UIs across platforms both costly and unreliable.

\noindent\textbf{Example.} 
We illustrate this challenge with the login page of \textit{OpenShop}~\cite{openshop}, a representative open-source Android app.
As shown in Figure~\ref{fig:motivation}, the interface combines multiple nested layouts (\texttt{RelativeLayout} and \texttt{LinearLayout}), text prompts, and styled buttons. 
To migrate such a design, developers must resolve three non-trivial issues:
(1) \emph{structural heterogeneity}, since Android XML expresses layouts imperatively while SwiftUI/Compose adopt declarative paradigms;
(2) \emph{resource incompatibility}, as color, dimension, and vector formats differ across platforms;
and (3) \emph{constraint simulation}, because constructs like \texttt{ConstraintLayout} have no direct counterpart in SwiftUI. 

\noindent\textbf{Motivation.} 
This example highlights the semantic gap between legacy Android XML and modern declarative frameworks.
Naïve syntactic translation cannot guarantee visual or behavioral fidelity, and existing alternatives either require significant manual effort or fail to preserve UI semantics. 
Therefore, an automated approach that operates on a \emph{semantic representation of UI structures and resources}, and systematically bridges the gap between XML and declarative paradigms, is needed. 
This motivates the design of \appname, which transpiles Android UIs into SwiftUI and Jetpack Compose while preserving both structural and visual semantics.

\section{Problem Definition}
\label{sec:problem-definition}
We model legacy Android UIs as structured artifacts and define the migration task as a \textit{semantics-preserving transformation} that bridges heterogeneous UI paradigms.

\textit{Input.}
An Android UI is denoted as $U_A = (L,V,R)$, where $L$ is the set of layout containers, $V$ the set of views, and $R$ the set of resources (e.g., colors, strings, images). 
Together, these elements encode both the structural hierarchy of the interface and the assets that ensure visual consistency.

\textit{Output.}
The goal is to produce a target declarative UI $U_T=(L',V',R')$ in Jetpack Compose or SwiftUI, which realizes the same layout intent and visual semantics while conforming to the idioms of the target framework.

\textit{UI Semantic Graph.}
We abstract $U_A$ into a graph $\mathcal{G}_A=(N_A,E_A)$, where nodes $N_A=L\cup V\cup R$ and edges $E_A$ capture three core relations:  
(i) containment $\mathsf{contains}(x,y)$, representing structural nesting;  
(ii) alignment and constraints $\mathsf{align}(x,y,\tau)$, representing spatial relations among elements; and  
(iii) resource dependencies $\mathsf{uses}(x,r)$, linking UI elements to external assets. The target UI $U_T$ is expressed as an analogous graph $\mathcal{G}_T$, enabling direct reasoning about structural and semantic correspondence.

\textit{Problem Statement.}
The migration task is to compute a mapping $h:\mathcal{G}_A\rightarrow \mathcal{G}_T$ and a resource adaptation $\rho:R\to R'$ that jointly satisfy:
\begin{enumerate}[label=\textbf{C\arabic*}, leftmargin=2.2em]
    \item \textbf{Structural preservation:} the hierarchical relations among layouts and views should be maintained or consistently approximated;
    \item \textbf{Visual fidelity:} the rendered appearance of the target UI should closely resemble the source, as measured by perceptual similarity metrics;
    \item \textbf{Semantic consistency:} essential attributes (e.g., size, color, text) should be adapted in a way that preserves their intended meaning.
\end{enumerate}

Due to heterogeneous syntaxes, non-isomorphic layouts (e.g., \texttt{ConstraintLayout}), and divergent resource formats, exact isomorphism is infeasible.  
Hence, the migration requires a semantic mapping strategy that interprets structural and visual intent at an intermediate level and reconstructs equivalent declarative constructs in the target framework.

%% file: approach.tex
\section{Proposed Approach}
\label{sec:approach}

\begin{figure*}
\centering
\includegraphics[width=0.95\linewidth]{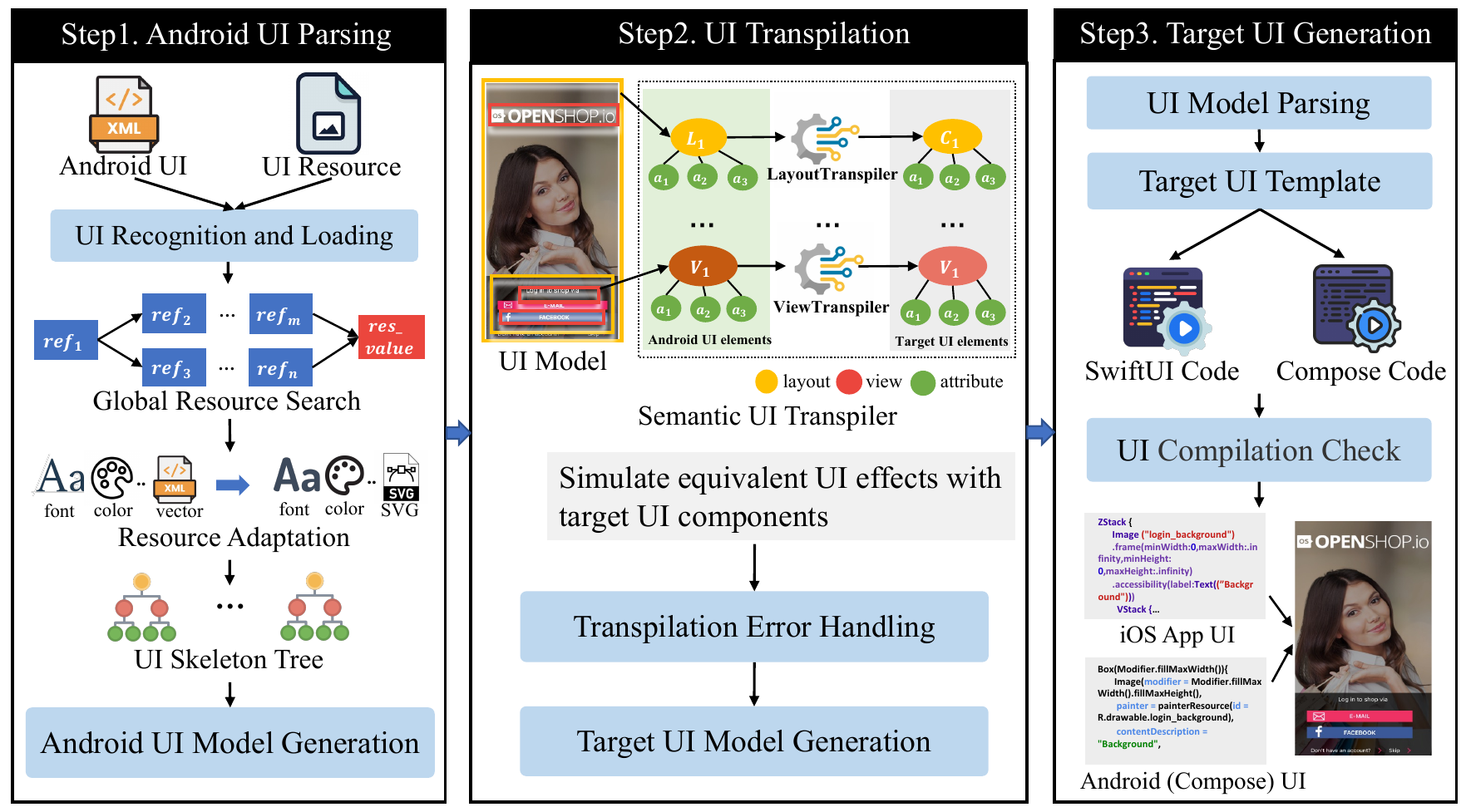}
\caption{Overview of the \appname pipeline.}
\label{fig:approach}
\end{figure*}

Figure~\ref{fig:approach} presents the overall framework of \appname. 
Starting from legacy Android XML UIs, our approach (i) parses layouts, views, and resources into a unified \emph{UI Semantic Graph}, (ii) applies semantic transpilation to map Android constructs into declarative counterparts, and (iii) generates compilable SwiftUI and Jetpack Compose code through platform-specific templates. 
This pipeline ensures that migration is both semantically faithful and practically executable.

\subsection{Android UI Parsing}
\label{sec:parsing}
The first step is to extract Android UI elements and represent them in a form suitable for semantic transformation. 
Rather than operating directly on XML syntax, we construct an intermediate abstraction that captures the structure and attributes of legacy UIs in a platform-independent manner, forming the basis of the UI Semantic Graph (Section~\ref{sec:problem-definition}).
Formally, we model an Android project as $U_A=(L,V,R)$, where $L$, $V$, and $R$ denote layouts, views, and resources defined under the \texttt{res} directory.
Dynamic components (e.g., \texttt{Activity} classes) are excluded, since they are written in Java/Kotlin and primarily encode business logic rather than declarative UI specifications.
Building on this abstraction, we define the UI Semantic Graph as $\mathcal{G}_A=(N_A,E_A)$, where nodes $N_A$ correspond to elements in $L \cup V \cup R$ and edges $E_A$ encode containment and reference relations.
This formalism provides a uniform representation that we use consistently in Section~\ref{sec:transpilation} to describe semantics-preserving transformations.

Parsing is challenging for three reasons.
First, resources often involve multiple layers of indirection (e.g., \texttt{@color/primary} → \texttt{@android:color/holo\_blue}).
Second, layouts rely on heterogeneous managers such as \texttt{LinearLayout}, \texttt{ConstraintLayout}, and \texttt{RelativeLayout}, which encode structure differently.
Third, attribute naming and value domains are inconsistent (e.g., dimensions may be expressed in \texttt{dp}, \texttt{px}, or \texttt{sp}).
Our parser therefore (a) extracts layouts and views into a hierarchical tree reflecting containment, (b) resolves resource references recursively through global lookup until concrete values are obtained, and (c) normalizes attributes (e.g., converting \texttt{16dp}, \texttt{16px}, and \texttt{16sp} into a uniform numeric representation). 
For example, given a snippet  
\texttt{<TextView android:textColor="@color/primary"/>},  
if \texttt{primary} is defined as \texttt{@android:color/holo\_blue}, our parser dereferences both layers and normalizes the final value into the canonical RGB representation.  

The output is a rooted forest where each node corresponds to a UI element $x$ with attributes $\alpha(x)$ and edges encode containment or resource usage. 
This forest is then embedded into the UI Semantic Graph $\mathcal{G}_A=(N_A,E_A)$, which provides a uniform semantic representation for downstream transpilation. 
Unlike traditional XML parsers, this semantic graph abstraction makes structural, visual, and resource information explicit, enabling systematic reasoning and facilitating generalization beyond Android-specific syntax.

\begin{table}[t]
\centering
\small
\caption{Structure of the Android \textit{res} directory and representative UI resource types. These resources are parsed and normalized as the first step of migration.}
\label{tab:res}
\rowcolors{2}{gray!10}{white}
\begin{tabularx}{\linewidth}{
  >{\bfseries}l
  >{\raggedright\arraybackslash}p{3.6cm}
  >{\raggedright\arraybackslash}X}
\toprule
Type & Example Files & Description \\
\midrule
Layout &
\path{res/layout/main.xml}, \path{res/layout/dashboard.xml} &
XML files that define screen layouts and hierarchical UI structures. \\

Drawable &
\path{res/drawable/icon.png}, \path{res/drawable/background.xml} &
Raster images and XML graphics used for icons and backgrounds. \\

String &
\path{res/values/strings.xml} &
Localized text resources displayed in the interface. \\

Color &
\path{res/values/colors.xml} &
Color definitions referenced by layouts and styles. \\

Dimension &
\path{res/values/dimens.xml} &
Values (e.g., margins, font sizes) ensuring consistent sizing across screens. \\

Style/Theme &
\path{res/values/styles.xml} &
Themes and style definitions that standardize visual appearance. \\

Vector &
\path{res/drawable/ic_launcher.xml} &
Scalable vector graphics for resolution-independent icons. \\
\bottomrule
\end{tabularx}
\end{table}

\subsubsection{UI Recognition and Loading}
Android projects follow a standardized structure in which user interface (UI) resources are organized under the \texttt{res} directory. 
As shown in Table~\ref{tab:res}, this directory contains layouts, drawables, strings, colors, dimensions, and styles that collectively define the static aspects of an app’s UI. 
Among them, layout XML files specify the hierarchical arrangement of views, while auxiliary resources (e.g., colors, dimensions, styles) provide reusable definitions that ensure visual consistency.  

To prepare for migration, we parse all XML layouts together with their referenced resources and represent them in a platform-independent form. 
Each layout is transformed into a hierarchical tree that captures view containment and attributes, while resource references are linked to their definitions rather than treated as raw symbols. 
This representation yields a complete inventory of the app’s static UI elements, abstracted away from Android-specific syntax, and serves as the entry point for constructing the UI Semantic Graph used in subsequent transpilation.

\subsubsection{UI Resource Normalization}
Android XML layouts frequently reference external resources such as colors, dimensions, or styles, which may themselves be defined through multiple layers of indirection. 
While this indirection improves modularity in Android, it complicates migration because declarative frameworks (e.g., Jetpack Compose, SwiftUI) expect explicit values instead of symbolic references. 
For example, \texttt{@color/textColor} may first resolve to \texttt{@color/account} before eventually mapping to a concrete hex value.  

To address this, we design a \emph{Global Resource Search (GRS)} mechanism that recursively resolves resource references until concrete values are obtained and substitutes them directly into the parsed UI tree. 
Starting from a resource reference (e.g., \texttt{@color/textColor}), GRS recursively follows indirections in Android resource tables (including \texttt{colors.xml}, \texttt{dimens.xml}, \texttt{strings.xml}, and drawable selectors) until a terminal value is reached. 
To ensure robustness, GRS maintains a \texttt{visited} set and enforces a maximum resolution depth to prevent cyclic dependencies and infinite expansion. If no terminal value can be resolved, the original reference is preserved and marked as unresolved for backend fallback handling. 
Through GRS, each UI element in the semantic UI graph is associated with an explicit, platform-agnostic representation of its attributes (e.g., normalized RGB color values or pixel-level dimensions).

Beyond resolution, Android and iOS also differ in resource formats and data models, making direct reuse infeasible. 
We therefore introduce a \emph{UI Resource Adapter} that maps normalized values into target-specific constructs.
Instead of directly mapping Android resources to platform APIs, the adapter applies \emph{typed adaptation rules} that interpret each resource according to its semantic role (e.g., color, dimension, text, or image) and translates it into a platform-agnostic representation.
These canonical forms (e.g., \emph{fill}, \emph{wrap}, or \emph{fixed} sizing semantics, resolved color values, or abstract image identifiers) are then instantiated by SwiftUI or Jetpack Compose.
In our implementation, the adapter covers four categories of adaptations (color, dimension, textual, and drawable resources), which account for the dominant resource usages observed in modern Android apps.
For example, Android hexadecimal colors are converted into SwiftUI’s \texttt{Color} initializers, while vector drawables are transformed into portable formats such as SVG or PDF. 
In Jetpack Compose migration, the same normalized values are injected into APIs like \texttt{colorResource} or \texttt{painterResource}.  
Together, GRS and the adapter implement a semantic normalization layer that consolidates heterogeneous Android resources into a uniform, cross-platform representation, enabling systematic and faithful transpilation to modern declarative frameworks.

In our implementation, the adapter includes four categories of adaptations—color, dimension, textual, and drawable resources—covering the dominant resource usages observed in modern Android apps.
Together, GRS and the adapter form a semantic normalization layer that consolidates heterogeneous Android resource definitions into a uniform cross-platform representation, enabling deterministic and consistent transpilation to modern declarative UI frameworks.

\subsubsection{Android UI Model Generation}
Android UIs consist of heterogeneous elements—layouts, views, and resources—connected by containment and reference relations. 
Accurately capturing these relationships is essential for migration but challenging due to nested hierarchies and indirect bindings.  
To this end, we first construct an Android UI Model—a structured tree rooted at the top-level layout, where internal nodes represent nested containers and leaves correspond to views (e.g., \texttt{Button}, \texttt{TextView}). 
Each node is annotated with resolved attributes and normalized resources, yielding a context-rich abstraction that captures both the structure and semantics of the original XML. 
This model is then lifted into the UI Semantic Graph, which generalizes Android-specific constructs into a platform-independent representation suitable for systematic reasoning and semantics-preserving transformation.

This model serves as the \emph{input} to the UI Semantic Graph. 
Concretely, the model captures the Android-specific hierarchy in a structured yet platform-dependent form, whereas the UI Semantic Graph generalizes it into a platform-independent graph representation that explicitly encodes structural, visual, and resource semantics. 
In other words, the model is a faithful semantic snapshot of an Android layout, and the graph lifts this snapshot into a uniform abstraction suitable for systematic reasoning and cross-platform transpilation.

\subsection{UI Transpilation}
\label{sec:transpilation}
The goal of this step is to transform the Android UI model into target declarative representations while preserving semantic fidelity. 
Direct syntactic translation from imperative XML to declarative frameworks such as SwiftUI or Jetpack Compose is infeasible: 
Android relies on nested imperative layouts (e.g., \texttt{RelativeLayout}, \texttt{ConstraintLayout}), whereas declarative frameworks describe composition through functional containers (e.g., \texttt{VStack}, \texttt{ZStack}, \texttt{Row}, \texttt{Box}). 
This structural mismatch, together with divergent resource systems, renders one-to-one mappings insufficient.

To address this, we design a \emph{Semantic UI Transpiler (SUT)} that operates at the semantic-level transformation. 
SUT takes the Android UI model (Section~\ref{sec:parsing}) as input and produces a target representation through three steps:
\ding{182} \textbf{Semantic interpretation.} Each layout or view is analyzed in terms of its functional role (e.g., sequential arrangement, overlapping composition, text rendering) rather than its literal XML tag.
\ding{183} \textbf{Semantic-to-declarative mapping.} A set of parametric transformation schemas maps these semantic roles into equivalent declarative constructs. 
For example, a horizontal \texttt{LinearLayout} corresponds to a \texttt{Row} in Compose or an \texttt{HStack} in SwiftUI, while a \texttt{ConstraintLayout} is decomposed into nested stacks augmented with alignment modifiers.
\ding{184} \textbf{Layout reconstruction.} The target layout is synthesized by composing transformed nodes while preserving containment, alignment, and resource bindings. 
This reconstruction step enforces both structural coherence and idiomatic usage of the target framework.

By reconstructing UIs from semantic roles instead of performing literal XML translations, SUT bridges heterogeneous UI paradigms and supports both cross-platform migration (Android $\to$ iOS) and intra-platform modernization (XML $\to$ Compose).
Rather than replicating Android-specific layouts, \appname preserves UI semantics and realizes them using the native declarative layout systems of the target platform.
Layout attributes such as \texttt{padding} and \texttt{spacing} are interpreted as relative layout semantics instead of fixed coordinates, and are therefore resolved by SwiftUI or Jetpack Compose through their device adaptive layout mechanisms.
This allows structural intent (e.g., containment, alignment, and spacing) to be preserved while naturally conforming to platform conventions such as screen-size adaptation, without enforcing platform-specific aesthetic redesign.
More broadly, SUT exemplifies a paradigm of \emph{semantics-guided graph transformation}, which can generalize to other software migration tasks where heterogeneous syntaxes must be reconciled through an intermediate semantic abstraction.

\begin{figure}
\centering
\includegraphics[width=0.95\linewidth]{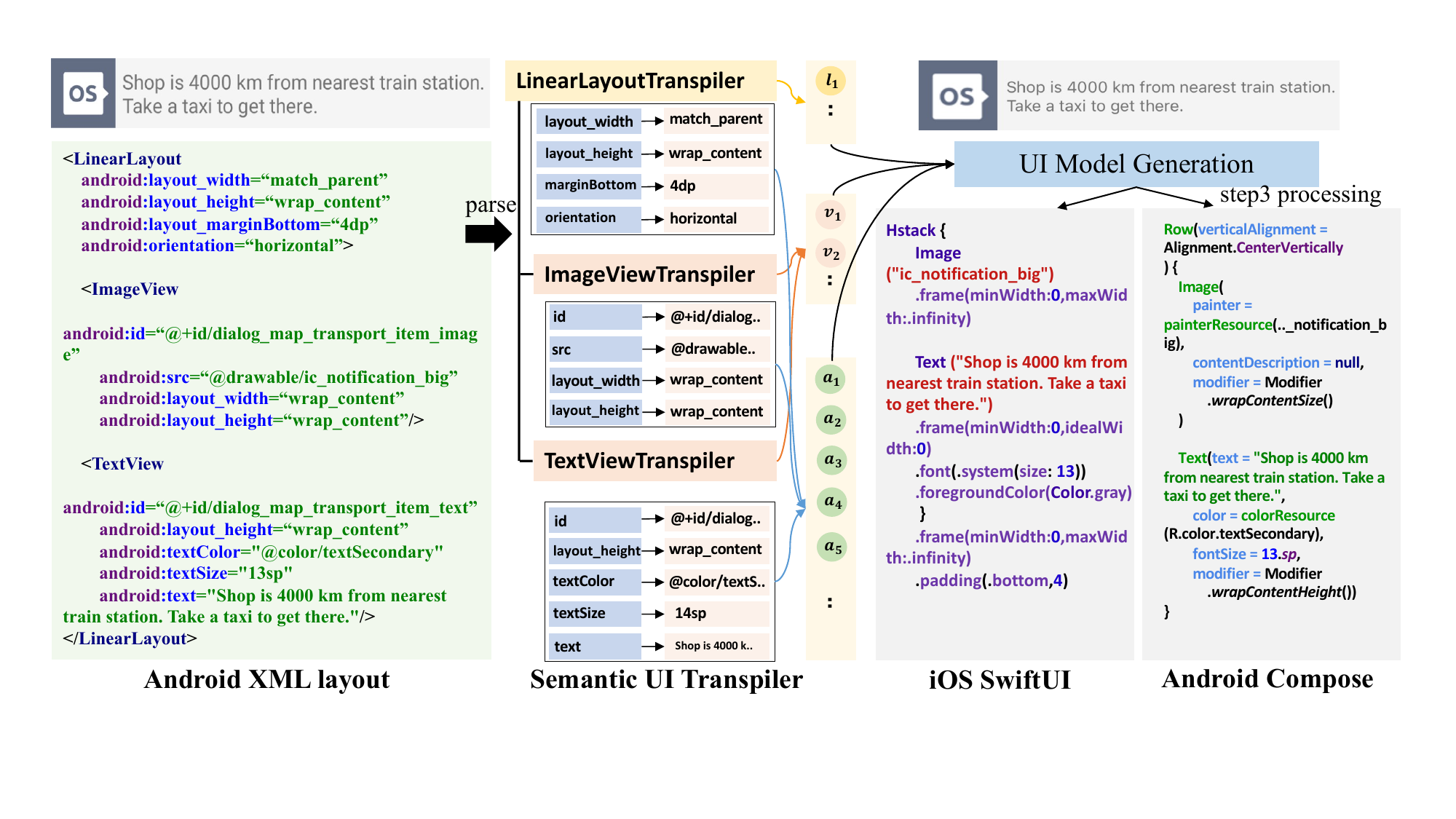}
\caption{An example of transpiling an Android UI layout to an equivalent visual interface using SUT. 
The figure shows how structural hierarchies and semantic roles in the source XML are interpreted and reconstructed into declarative constructs, preserving layout intent while adapting to the target platform.
}
\label{fig:sut}
\vspace{-0.2cm}
\end{figure}

\subsubsection{UI Transpilation Process}
As illustrated in Figure~\ref{fig:approach}, the \emph{Semantic UI Transpiler (SUT)} converts the Android UI model into target declarative structures through a semantics-guided graph transformation. 
Each XML layout is first represented as a rooted tree, with layout containers as internal nodes and views as leaves. 
SUT traverses this tree and interprets each node in terms of its semantic role (e.g., sequential arrangement, overlay, relative positioning), before reconstructing an equivalent declarative tree in SwiftUI or Jetpack Compose.

For simple layouts, the mapping is direct: a horizontal \texttt{LinearLayout} corresponds to \texttt{HStack} (SwiftUI) or \texttt{Row} (Compose), while a vertical one maps to \texttt{VStack} or \texttt{Column}. 
View components are similarly mapped, such as \texttt{ImageView}$\to$\texttt{Image} and \texttt{TextView}$\to$\texttt{Text}. 
Because both SwiftUI and Compose share declarative paradigms, SUT applies the same semantic mappings across platforms, with only minor syntactic adaptations.
The main challenge lies in constructs like \texttt{ConstraintLayout}, which encode relational constraints that are neither directly supported in SwiftUI nor idiomatic in modern Compose. 
SUT handles complex layouts such as \texttt{ConstraintLayout} by \emph{semantic decomposition}. Instead of enumerating low-level constraints, SUT abstracts them into relative positioning semantics (e.g., align-to-parent, align-to-sibling). 
These semantics are then systematically realized via nested declarative containers (\texttt{ZStack}/\texttt{Box}, \texttt{VStack}/\texttt{Column}, \texttt{HStack}/\texttt{Row}) with alignment modifiers, padding, and spacers. 
This abstraction layer ensures that even intricate constraints are faithfully reconstructed in an idiomatic style.

\begin{figure}
\centering
\includegraphics[width=0.9\linewidth]{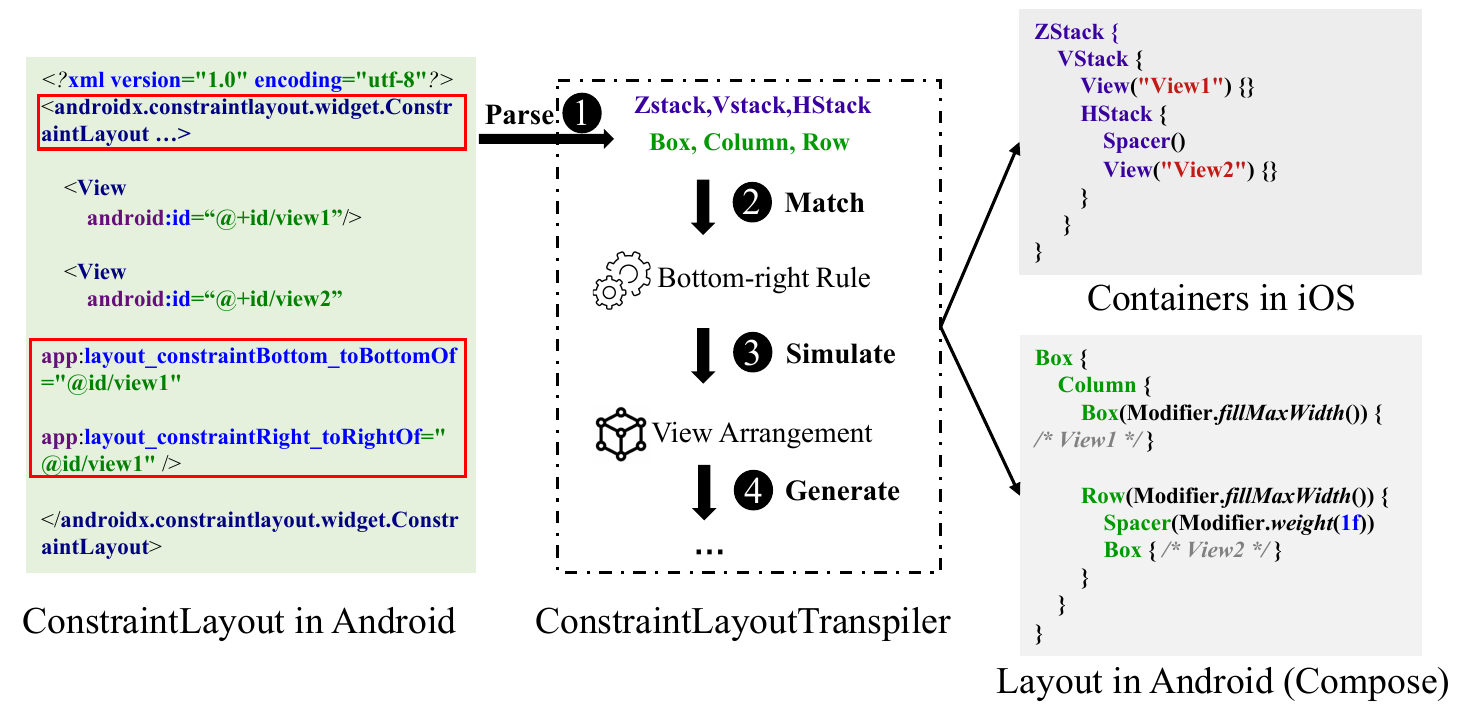}
\caption{
Transpilation process of converting Android’s \texttt{ConstraintLayout} to visually equivalent container combinations using SUT. 
}
\label{fig:constraintlayout}
\vspace{-0.3cm}
\end{figure}

As illustrated in Figure~\ref{fig:constraintlayout}, an Android \texttt{ConstraintLayout} may specify relative positioning rules such as \texttt{layout\_constraintBottom\_toBottomOf} or \texttt{layout\_constraintStart\_toStartOf}. 
Rather than directly encoding these constraints, SUT decomposes them into equivalent declarative structures: for SwiftUI, constraints are mapped to compositions of \texttt{ZStack}, \texttt{VStack}, and \texttt{HStack} with explicit alignment and spacing; for Jetpack Compose, analogous effects are achieved with \texttt{Row}, \texttt{Column}, and \texttt{Spacer}.  
This decomposition preserves the original relative positioning semantics (e.g., \textit{align to parent bottom}, \textit{center horizontally within parent}) while producing idiomatic target code that aligns with the design philosophy of declarative frameworks. 
By eliminating low-level constraint specifications and replacing them with higher-level semantic containers, the generated code is not only visually faithful but also more maintainable and extensible in the target platform.

\textit{Extensibility.} 
Real-world Android \apps typically rely on a recurring set of standard layouts and views. 
SUT currently supports 30 component types and 75 attributes, which already cover the majority of practical cases. 
To remain adaptable to evolving UI frameworks, SUT adopts a \emph{semantic plugin architecture}: new components are added by declaring their semantic roles and associating them with target constructs. 
This modular design ensures that extending support to additional layouts or attributes requires minimal effort and preserves maintainability over time.

\textit{Error Handling.} 
In practice, migration often encounters components or attributes without direct counterparts on the target platform. 
SUT ensures robustness through a two-level fallback strategy: (i) unrecognized layouts are replaced with generic containers that preserve structural relations, and (ii) unknown views are substituted by placeholders annotated with their key attributes (e.g., size, position, color). 
This guarantees syntactic validity and preserves high-level visual intent, even in the absence of exact mappings. 
In addition, SUT generates a detailed migration report that records unresolved elements and the reasons for failure, enabling developers to refine transpilation or integrate custom plugins. 
This mechanism not only accommodates third-party components but also illustrates a broader methodological principle: \emph{semantic degradation with traceability}, where missing constructs are gracefully approximated while maintaining developer visibility and control.

\subsubsection{Target UI Model Generation}
The output of SUT is a structured representation of the target UI, referred to as the \emph{Target UI Model}. 
Each model encodes container types (e.g., \texttt{ZStack}, \texttt{VStack}), associated attributes, and nested views, forming a hierarchical abstraction of the interface. 
This intermediate representation is platform-neutral: it suppresses syntactic details while retaining all semantic information needed for code realization. 
By aggregating such models, we obtain a complete, declarative description of the target application’s UI.

\subsection{Target UI Generation}
The final step translates Target UI Models into executable SwiftUI or Jetpack Compose code. 
The objective is not only to produce compilable source files but also to ensure that the code is \emph{idiomatic}, preserving both semantic fidelity and target-specific conventions.

\subsubsection{UI Model Parsing}
Before code emission, we perform platform-specific validation on the Target UI Models. 
Declarative frameworks impose subtle ordering and scoping rules that directly impact rendering and compilation.  
For SwiftUI, for example, modifier ordering is crucial: \texttt{frame} must precede \texttt{background} to ensure correct sizing, while \texttt{padding} must be placed before alignment modifiers to avoid layout inconsistencies. 
Jetpack Compose enforces similar rules through chained modifier functions, where improper ordering can lead to visually incorrect layouts.

To address this, each Target UI Model is augmented with an ordered modifier list derived from dependency constraints between attributes. 
For instance, dimension-related attributes are resolved before styling attributes, and spacing must be placed before alignment. 
This ordering is enforced during code generation, guaranteeing that the emitted code is both syntactically valid and visually consistent.  
The validated models thus form a reliable basis for synthesis into executable UI files.

\subsubsection{UI Code Generation}
\begin{wrapfigure}{r}{0.5\textwidth}
\centering
\includegraphics[width=\linewidth]{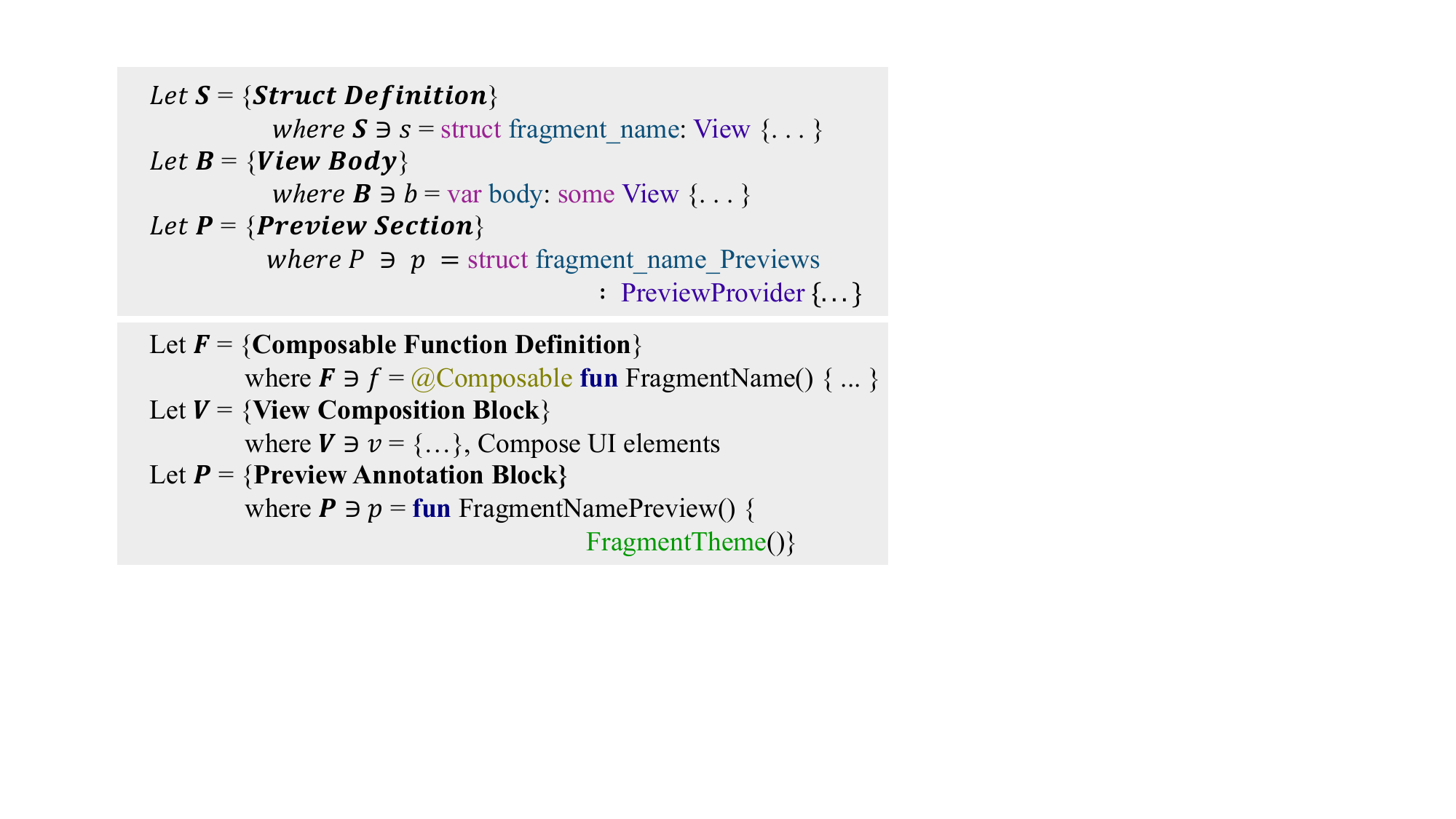}
\caption{SwiftUI and Compose code generation templates. 
They define how Target UI Models are instantiated into platform-specific constructs (e.g., \texttt{struct}/\texttt{@Composable}), ensuring the generated code is both compilable and idiomatic.}
\label{fig:ui_generate}
\end{wrapfigure}

Code generation is template-driven but semantically informed. 
Each Target UI Model node is mapped to a code template parameterized by its attributes and child elements.  
As shown in Figure~\ref{fig:ui_generate}, for SwiftUI, templates generate struct definitions with a \texttt{body} function and optional \texttt{PreviewProvider} constructs to enable live rendering in Xcode. 
For Jetpack Compose, the system emits Kotlin \texttt{@Composable} functions annotated with \texttt{@Preview}, ensuring direct integration with Android Studio’s preview tooling.

Beyond direct translation, SUT also injects target-specific idioms to improve readability and maintainability.  
For example, groups of aligned child views are wrapped in higher-order containers (\texttt{Group} in SwiftUI or \texttt{Column}/\texttt{Row} in Compose), and reusable resources are factored into constants to avoid duplication.  
We also generate scaffolding code (e.g., \texttt{import SwiftUI}, \texttt{setContent} in Compose) so that the synthesized output is self-contained and immediately compilable.  

This design balances automation with human-centric readability: the generated code is not only functionally equivalent but also stylistically aligned with the conventions of SwiftUI and Compose developers, facilitating downstream modification and reuse.

\subsubsection{UI Compilation Check}
Finally, we validate the generated code through native toolchains to ensure practical usability.
SwiftUI code is compiled in Xcode and inspected via its live preview engine, while Compose code is compiled in Android Studio and rendered with Compose Preview.
When compilation or rendering fails, \appname produces an automated migration report that pinpoints the responsible UI elements and attributes, and provides both the failing code and suggested fallback constructs.
This mechanism prevents silent failures and shortens the refinement loop for developers, turning migration results into actionable outputs.
In this way, \appname guarantees that the final code is not only semantically faithful but also executable within production environments, completing the end-to-end migration pipeline from Android XML to modern declarative frameworks.

%% file: evaluation.tex
\section{Evaluation}
\label{sec:evaluation}
We evaluate \appname through a series of experiments designed to answer the following research questions:

\begin{itemize}[leftmargin=*]
\item \textbf{RQ1: Migration Quality.} How is the quality of the UI migrated by \appname? 
\item \textbf{RQ2: Migration Effectiveness.} How effective is \appname in preserving UI fidelity and reducing developer effort?
\item \textbf{RQ3: Development Effort Reduction.} How does \appname reduce manual development effort in UI transpilation?
\item \textbf{RQ4: Migration Performance.} What is the runtime efficiency of \appname in UI migration?
\end{itemize}

\subsection{Experimental Setup}
\textit{Dataset.} 
Since UI migration requires full access to application resources, we draw candidate projects from the \texttt{open-source-android-apps} collection on GitHub~\cite{openapps}, which aggregates widely used open-source Android applications (9.8k stars). 
To ensure both reliability and diversity, we apply the following criteria:  
(i) the project must provide complete source code and use \texttt{Gradle} as its build system, ensuring a consistent project structure for resource parsing; and  
(ii) the project should represent a broad range of domains, allowing us to evaluate the generality of migration across heterogeneous UI designs.  

From this pool, we randomly sample 40 projects and verify that they can be successfully built, compiled, and previewed. 
After excluding those that fail to meet these requirements, we retain 31 applications spanning ten domains (e.g., business, communication, media, social networking). 
As summarized in Table~\ref{tab:dataset}, the dataset covers 500 XML files comprising 1,027 layouts and 2,583 views, totaling 37,417 lines of UI code. 
This collection captures both simple and complex UI patterns, including nested layouts and resource-heavy designs, providing a realistic benchmark for assessing semantic migration.  
To promote reproducibility, the complete dataset list is publicly available at~\cite{uimigratorDatasets}. 
This dataset serves as the foundation for our evaluation of migration completeness, semantic fidelity, and execution robustness in subsequent experiments.

\begin{table}[t]
\centering
\small
\caption{Overview of the Android UI migration dataset across ten domains. 
We report the number of XML files, layouts, views, and lines of code (LOC), as well as average layouts per XML and views per layout to indicate structural complexity.}
\rowcolors{2}{gray!10}{white}
\begin{tabular}{lrrrrrr}
\toprule
\textbf{Domain} & \textbf{XMLs} & \textbf{Layouts} & \textbf{Views} & \textbf{LOC} & \textbf{Layouts/XML} & \textbf{Views/Layout} \\
\midrule
Business       &  57  & 188  & 302  &  4,740 & 3.30 & 1.61 \\
Communication  &  41  &  84  & 214  &  4,029 & 2.05 & 2.55 \\
Education      &  27  &  36  & 132  &  1,893 & 1.33 & 3.67 \\
Finance        &  68  & 128  & 376  &  5,281 & 1.88 & 2.94 \\
Health         &  13  &  28  & 143  &  1,825 & 2.15 & 5.11 \\
Media          &  35  &  57  & 121  &  2,505 & 1.63 & 2.12 \\
News           &  69  & 128  & 193  &  2,419 & 1.86 & 1.51 \\
Productivity   &  29  &  51  &  97  &  1,809 & 1.76 & 1.90 \\
Network         &  44  &  96  & 151  &  2,304 & 2.18 & 1.57 \\
Tools          & 117  & 231  & 854  & 10,612 & 1.97 & 3.70 \\
\midrule
\rowcolor{gray!20}
\textbf{Total} & \textbf{500} & \textbf{1,027} & \textbf{2,583} & \textbf{37,417} & \textbf{2.05} & \textbf{2.52} \\
\bottomrule
\end{tabular}
\label{tab:dataset}
\end{table}

\begin{figure}
   \centering
   \begin{mdframed}[backgroundcolor=lightgreen!30, linecolor=gray, outerlinewidth=1pt]
    \textbf{Role:} \textit{You are an expert in the field of Android and iOS, specializing in equivalent UI transpilation}.  \\
    \textbf{User:} \textit{Given the following \textcolor{blue}{Android XML code} for a GUI layout, transpile it into an equivalent \textcolor{blue}{SwiftUI or Compose code} that achieves the \textcolor{blue}{same UI effect}.}  \\
    \textbf{Original UI:} \texttt{\{Android app UI\}}
   \end{mdframed}
   \caption{Prompt Template for transpiling Android UI to declarative frameworks (SwiftUI and Jetpack Compose).}
   \label{fig:ui-prompt-example}
\end{figure}

\textit{Baseline}.
To the best of our knowledge, \appname is the first approach that automates end-to-end migration from Android XML layouts to declarative UI frameworks (SwiftUI and Jetpack Compose). 
We are not aware of any prior tool that directly supports this task. 
Therefore, we use the latest proprietary ChatGPT model (GPT-5.2) and the open-source DeepSeek-V3.2 as baselines, both accessed with default parameters and without fine-tuning.
We design a unified prompt that supplies the source Android XML and explicitly requests translation into either SwiftUI or Jetpack Compose, including the target platform and expected output format (Figure~\ref{fig:ui-prompt-example}). 
To ensure fairness, we avoid task-specific tool augmentation or bespoke post-processing, limiting the interaction to what an end user could reasonably provide in a single prompt.  

We do not include screenshot-to-UI generation methods (e.g., \cite{beltramelli2018pix2code}) in the main quantitative benchmark, as they address a related but different problem formulation.
Such methods take images as input and attempt to reconstruct UI code, focusing primarily on visual similarity.
In contrast, our task is \textit{semantics-preserving migration}: the input already contains a complete XML specification with explicit layout hierarchies, resource bindings, and constraints.
Migration therefore requires faithfully reusing this semantic information and mapping it into idiomatic target constructs, rather than inferring structure from pixels.
Because the inputs, available information, and evaluation criteria differ substantially, a direct head-to-head comparison in the main evaluation would be inappropriate.
Instead, we provide an exploratory comparison in the Discussion~\ref{discussion_exploratory} to examine their complementary strengths.

\begin{table*}[t]
\centering
\footnotesize
\setlength{\tabcolsep}{3pt}
\renewcommand{\arraystretch}{0.95}
\rowcolors{12}{gray!9}{white}
\caption{Results of UI migration across ten domains.
For each category, we report the number of successfully migrated elements (M),
the number of omitted elements (O), and the migration rate (\%).
\textbf{Err.} counts syntax errors in generated code.}
\label{tab:migration}
\resizebox{\textwidth}{!}{%
\begin{tabular}{l
ccc ccc ccc c|
ccc ccc ccc c}
\toprule
& \multicolumn{10}{c}{\textbf{Migration to Jetpack Compose}}
& \multicolumn{10}{c}{\textbf{Migration to SwiftUI}} \\
\cmidrule(lr){2-11}\cmidrule(lr){12-21}

\textbf{Domain}
& \multicolumn{3}{c}{\textbf{XML}}
& \multicolumn{3}{c}{\textbf{Layouts}}
& \multicolumn{3}{c}{\textbf{Views}}
& \textbf{Err.}
& \multicolumn{3}{c}{\textbf{XML}}
& \multicolumn{3}{c}{\textbf{Layouts}}
& \multicolumn{3}{c}{\textbf{Views}}
& \textbf{Err.} \\
\cmidrule(lr){2-4}\cmidrule(lr){5-7}\cmidrule(lr){8-10}
\cmidrule(lr){12-14}\cmidrule(lr){15-17}\cmidrule(lr){18-20}

& M & O & (\%) & M & O & (\%) & M & O & (\%) & 
& M & O & (\%) & M & O & (\%) & M & O & (\%) & \\
\midrule
Business      & 53 &  4 &  93 & 185 &  3 &  98 & 295 &  7 &  98 &  3
              & 54 &  3 &  94 & 186 &  2 &  98 & 297 &  5 &  98 &  5 \\
Comm.         & 30 &  11 &  73 &  68 & 16 &  81 & 153 & 61 &  71 &  6
              & 30 &  11 &  73 &  68 & 16 &  81 & 153 & 61 &  71 &  8 \\
Education     & 25 &  2 &  92 &  31 &  5 &  86 & 121 & 11 &  92 &  1
              & 25 &  2 &  92 &  32 &  4 &  88 & 125 &  7 &  94 &  2 \\
Finance       & 60 &  8 &  88 & 119 &  9 &  93 & 347 & 29 &  92 &  8
              & 56 &  12 &  82 & 113 & 15 &  88 & 310 & 66 &  84 & 11 \\
Health        & 13 &  0 & 100 &  28 &  0 & 100 & 143 &  0 & 100 &  0
              & 13 &  0 & 100 &  28 &  0 & 100 & 143 &  0 & 100 &  0 \\
Media         & 32 &  3 &  91 &  55 &  2 &  96 & 115 &  6 &  95 &  1
              & 31 &  4 &  88 &  54 &  3 &  94 & 111 & 10 &  91 &  2 \\
News          & 67 &  2 &  97 & 126 &  2 &  98 & 188 &  5 &  97 &  4
              & 67 &  2 &  97 & 125 &  3 &  97 & 186 &  7 &  96 &  3 \\
Productivity  & 27 &  2 &  93 &  49 &  2 &  96 &  92 &  5 &  94 &  7
              & 26 &  3 &  89 &  48 &  3 &  94 &  87 & 10 &  89 &  9 \\
Network       & 43 &  1 &  97 &  92 &  4 &  96 & 150 &  1 &  99 &  5
              & 42 &  2 &  95 &  90 &  6 &  93 & 149 &  2 &  98 &  2 \\
Tools         &110 &  7 &  94 & 228 &  3 &  98 & 810 & 44 &  94 &  8
              &109 &  8 &  93 & 226 &  5 &  97 & 767 & 87 &  89 & 10 \\
\midrule
\textbf{Total}&460 & 40 & 92.0 & 981 & 46 & 95.5 & 2414 & 169 & 93.5 & 43
              &453 & 47 & 90.6 & 970 & 57 & 94.5 & 2328 & 255 & 90.1 & 61 \\
\bottomrule
\end{tabular}}
\end{table*}

\subsection{RQ1: How is the quality of the UI migrated by \appname?}
We evaluate migration quality using two complementary metrics: 
(1) \emph{migration completeness}, i.e., the proportion of UI elements successfully transpiled; and 
(2) \emph{migration correctness}, i.e., whether the generated code compiles and renders the intended layout without breaking semantics.

\begin{figure}
\centering
\includegraphics[width=\linewidth]{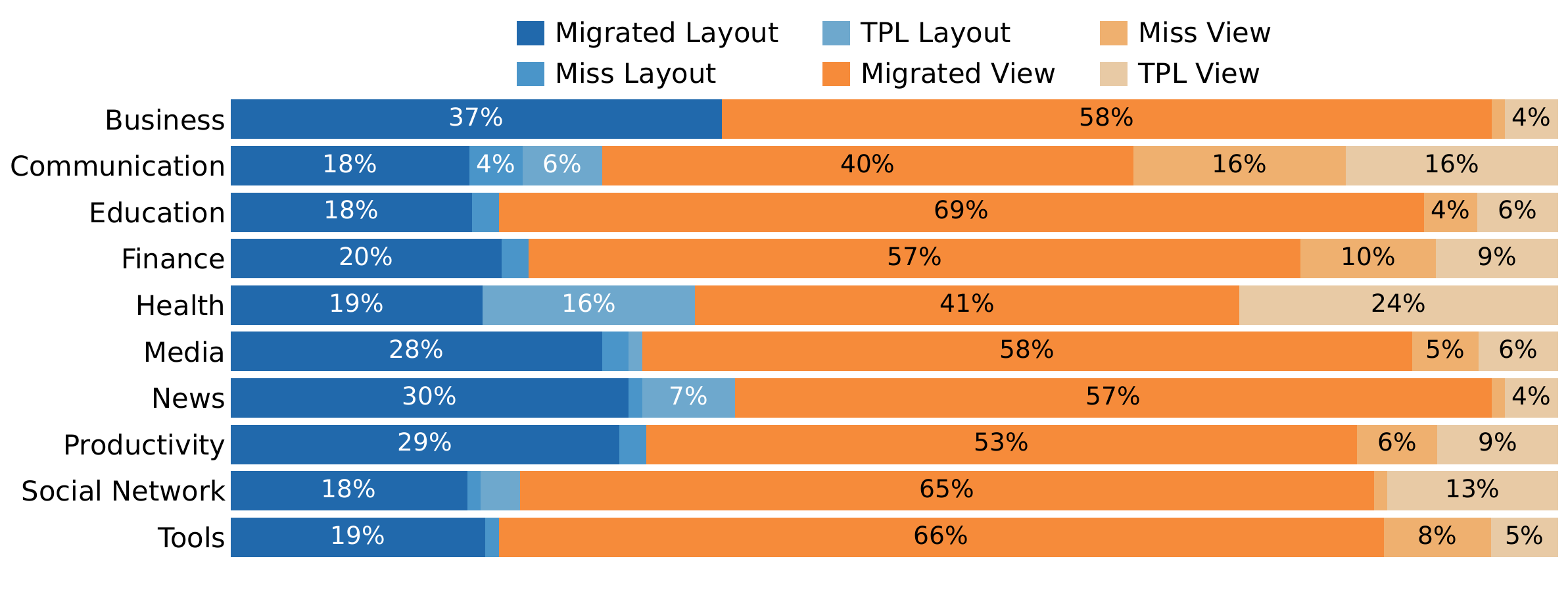}
\caption{Proportion of layouts and views migrated by \appname within an app, as well as the proportion of third-party library UI layouts and view types within the app.}
\label{fig:ui_migration}
\end{figure}

\textit{Migration Completeness.} 
We define completeness as the \emph{migration rate}, the ratio of migrated UI elements (XML files, layouts, and views) to the total number of elements in the original Android project (Table~\ref{tab:dataset}). 
As shown in Table~\ref{tab:migration}, \appname consistently achieves high coverage across all ten domains. 
For Jetpack Compose, migration rates reach 92.0\%, 95.5\%, and 93.5\% for XML files, layouts, and views, respectively. 
For SwiftUI, the corresponding rates are 90.6\%, 94.5\%, and 90.1\%. 
The slightly lower SwiftUI coverage reflects its greater structural mismatch with Android XML. 
Per-domain analysis (Figure~\ref{fig:ui_migration}) further shows that domains dominated by standard layouts (e.g., health) achieve near-complete migration, while those relying heavily on custom widgets (e.g., business) exhibit lower coverage.
Manual inspection shows that these omissions fall into three categories: (i) third-party or proprietary widgets without declarative counterparts,(ii) missing or dynamically generated resources, and (iii) non-standard XML constructs that violate layout specifications.
These cases are explicitly recorded in the migration report and typically require only localized adjustments.

\textit{Migration Correctness.} 
Executability Correctness is assessed by compiling the generated code and validating its rendering in native preview tools. 
We recorded 43 syntax errors in Compose and 61 in SwiftUI. 
Roughly 70\% of these errors are due to unresolved external resources (e.g., missing image assets), while the remainder involve unsupported components such as \texttt{MultiAutoCompleteTextView}, which lacks a declarative counterpart. 
In such cases, placeholders are inserted and issues logged in a migration report to guide developer refinement.  

Overall, \appname achieves migration rates above 90\% on both targets, and the majority of errors are localized to unsupported or missing resources rather than systematic failures. 
These results indicate that \appname preserves most UI structures across diverse domains and produces compilable, visually faithful code with minimal manual intervention. 
This provides strong evidence that semantics-preserving transpilation is both practical and effective for cross-platform UI migration.

\subsection{RQ2: How effective is \appname in preserving UI fidelity and reducing developer effort?}
A central goal of migration is to preserve the visual consistency of the original UI while minimizing the amount of manual refinement required by developers. 
Accordingly, we evaluate effectiveness along two dimensions: 
(1) \emph{visual fidelity}, measured by the Structural Similarity Index (SSIM) between screenshots of original and migrated previews.
To complement this pixel-level measure, we introduce \emph{Project-wide Semantic Coherence (PSC)}, which evaluates whether the migrated UI preserves key structural semantics beyond visual similarity.
PSC compares the source and migrated interfaces on their normalized semantic graphs along three dimensions:
(S) consistency of component hierarchy,
(A) preservation of normalized key attributes (e.g., text, size, color, and resource bindings), and
(L) retention of essential layout relations such as containment, alignment, and proportional spacing.
We compute $\emph{PSC} = \frac{S + A + L}{3}$; and (2) \emph{manual adjustment effort}, quantified by Code Relative Change (CRC) and Code Token Change (CTC), which capture the extent of post-migration edits needed for compilability and visual correctness.

\begin{table*}[t]
\centering
\footnotesize
\setlength{\tabcolsep}{4pt}
\renewcommand{\arraystretch}{1.05}
\rowcolors{12}{gray!9}{white}
\caption{Migration effort across ten domains.
CRC and CTC quantify post-migration code adjustment cost; lower values are better.}
\label{tab:effort_crc_ctc}
\resizebox{\textwidth}{!}{%
\begin{tabular}{l
cc cc cc|
cc cc cc}
\toprule
& \multicolumn{6}{c}{\textbf{Migration to Jetpack Compose}} & \multicolumn{6}{c}{\textbf{Migration to SwiftUI}} \\
\cmidrule(lr){2-7}\cmidrule(lr){8-13}

\textbf{Domain}
& \multicolumn{2}{c}{\textbf{ChatGPT-5.2}}
& \multicolumn{2}{c}{\textbf{DeepSeek-V3.2}}
& \multicolumn{2}{c}{\textbf{\appname}}
& \multicolumn{2}{c}{\textbf{ChatGPT-5.2}}
& \multicolumn{2}{c}{\textbf{DeepSeek-V3.2}}
& \multicolumn{2}{c}{\textbf{\appname}} \\
\cmidrule(lr){2-3}\cmidrule(lr){4-5}\cmidrule(lr){6-7}
\cmidrule(lr){8-9}\cmidrule(lr){10-11}\cmidrule(lr){12-13}
& CRC & CTC & CRC & CTC & CRC & CTC
& CRC & CTC & CRC & CTC & CRC & CTC \\
\midrule
Business   & 2.71 & 0.58 & 2.93 & 0.64 & 2.32 & 0.39 & 2.89 & 0.52 & 3.08 & 0.59 & 2.41 & 0.18 \\
Comm.      & 2.64 & 0.51 & 2.88 & 0.57 & 2.90 & 0.40 & 1.48 & 0.31 & 1.66 & 0.36 & 4.72 & 0.21 \\
Edu.       & 4.38 & 0.33 & 4.61 & 0.39 & 3.85 & 0.26 & 5.12 & 0.24 & 5.36 & 0.29 & 4.08 & 0.19 \\
Finance    & 2.44 & 0.27 & 2.63 & 0.31 & 2.86 & 0.16 & 3.21 & 0.30 & 3.39 & 0.35 & 2.19 & 0.11 \\
Health     & 2.31 & 0.61 & 2.55 & 0.68 & 1.82 & 0.37 & 1.96 & 0.55 & 2.14 & 0.63 & 1.91 & 0.10 \\
Media      & 2.73 & 0.36 & 2.95 & 0.41 & 2.49 & 0.18 & 2.58 & 0.29 & 2.77 & 0.33 & 1.96 & 0.07 \\
News       & 1.94 & 0.40 & 2.16 & 0.46 & 2.31 & 0.30 & 1.77 & 0.33 & 1.95 & 0.37 & 2.46 & 0.09 \\
Prod.      & 3.62 & 2.61 & 3.81 & 2.74 & 4.08 & 2.54 & 3.55 & 2.68 & 3.72 & 2.82 & 3.97 & 2.49 \\
Network    & 2.22 & 0.19 & 2.39 & 0.23 & 1.92 & 0.12 & 1.98 & 0.15 & 2.14 & 0.19 & 1.70 & 0.07 \\
Tools      & 1.82 & 0.10 & 2.01 & 0.13 & 1.26 & 0.05 & 1.63 & 0.09 & 1.79 & 0.12 & 0.99 & 0.03 \\

\midrule
\textbf{Avg} & 2.68 & 0.56 & 2.89 & 0.63 & 2.58 & 0.38 & 2.72 & 0.55 & 2.90 & 0.61 & 2.47 & 0.34 \\
\bottomrule
\end{tabular}}
\end{table*}

\textit{Manual Adjustment Effort.} 
For code-level corrections, \appname achieves CRC/CTC values of 2.88\%/0.50\% on Compose and 2.85\%/0.32\% on SwiftUI, meaning that over 97\% of the generated code is directly usable without modification. 
The LLM baselines (ChatGPT-5.2 and DeepSeek-V3.2) yield numerically comparable CRC/CTC values, indicating that they can often produce syntactically plausible UI code.
However, these aggregate metrics mask a key difference in error characteristics: LLM-generated outputs often require repeated localized adjustments to restore cross-file semantic consistency, which accumulates as post-migration effort.
By contrast, \appname integrates platform-specific constraints into transpilation, ensuring that even if placeholders are needed, the output remains compilable and extensible.

\textit{Visual Fidelity.} 
To further understand migration fidelity, we additionally analyze both visual and semantic preservation using complementary metrics.
Residual mismatches largely stem from (i) platform-specific style differences between Android and iOS, (ii) device-level variations in resolution, and (iii) unsupported third-party components replaced by placeholders. 
These cases are systematically logged in migration reports, enabling developers to selectively refine critical components.

Table~\ref{tab:quality_ssim_psc} reports SSIM and PSC, which capture complementary aspects of migration fidelity.
LLM-based approaches achieve SSIM scores comparable to \appname\ (Compose: 73.1\% for ChatGPT-5.2 and 72.0\% for DeepSeek-V3.2 vs.\ 81.9\% for \appname; 
SwiftUI: 66.1\% and 64.9\% vs.\ 78.2\%), indicating that modern models can generate visually plausible layouts.
However, their PSC scores are consistently lower (Compose: 73.1\% and 72.0\% vs.\ 93.2\% for \appname) and show greater variation across domains.
This difference arises because LLMs synthesize UI code in a file-local manner without enforcing a shared semantic representation, which can lead to inconsistencies in layout hierarchy, resource binding, and component relationships.
In contrast, \appname derives target code from an explicit semantic intermediate representation, enabling preservation of structural and cross-component semantics throughout the migration process.

Overall, \appname consistently delivers higher visual fidelity and similar or lower adjustment effort. 
These results demonstrate that semantics-guided transpilation provides a more reliable and developer-friendly foundation for cross-platform UI migration than ad-hoc LLM prompting.

\begin{table*}[t]
\centering
\footnotesize
\setlength{\tabcolsep}{4pt}
\renewcommand{\arraystretch}{1.05}
\rowcolors{3}{gray!9}{white}
\caption{Semantic preservation quality across ten domains.
SSIM measures visual similarity, while PSC evaluates project-level structural consistency. Higher values are better.}
\label{tab:quality_ssim_psc}
\resizebox{\textwidth}{!}{%
\begin{tabular}{l
cc cc cc|
cc cc cc}
\toprule
& \multicolumn{6}{c}{\textbf{Migration to Jetpack Compose}} & \multicolumn{6}{c}{\textbf{Migration to SwiftUI}} \\
\cmidrule(lr){2-7}\cmidrule(lr){8-13}

\textbf{Domain}
& \multicolumn{2}{c}{\textbf{ChatGPT-5.2}}
& \multicolumn{2}{c}{\textbf{DeepSeek-V3.2}}
& \multicolumn{2}{c}{\textbf{\appname}}
& \multicolumn{2}{c}{\textbf{ChatGPT-5.2}}
& \multicolumn{2}{c}{\textbf{DeepSeek-V3.2}}
& \multicolumn{2}{c}{\textbf{\appname}} \\

\cmidrule(lr){2-3}\cmidrule(lr){4-5}\cmidrule(lr){6-7}
\cmidrule(lr){8-9}\cmidrule(lr){10-11}\cmidrule(lr){12-13}
& SSIM & PSC & SSIM & PSC & SSIM & PSC
& SSIM & PSC & SSIM & PSC & SSIM & PSC \\

\midrule
Business & 72.6 & 71.2 & 71.4 & 68.4 & 74.5 & 91.5 & 64.1 & 69.3 & 63.3 & 66.1 & 68.4 & 88.2 \\
Comm.    & 72.9 & 70.6 & 72.0 & 67.9 & 73.8 & 89.8 & 70.2 & 68.2 & 69.0 & 65.5 & 72.9 & 86.7 \\
Edu.     & 70.1 & 69.7 & 69.2 & 66.9 & 74.1 & 92.4 & 53.8 & 63.4 & 52.7 & 60.8 & 73.2 & 90.3 \\
Finance  & 78.0 & 73.5 & 76.9 & 70.2 & 80.7 & 93.1 & 68.9 & 70.4 & 67.6 & 67.8 & 80.6 & 91.6 \\
Health   & 67.4 & 75.1 & 66.3 & 72.4 & 78.5 & 95.2 & 64.5 & 72.9 & 63.4 & 70.5 & 77.6 & 93.8 \\
Media    & 56.2 & 64.8 & 55.1 & 61.9 & 78.5 & 90.6 & 50.7 & 60.7 & 49.6 & 57.3 & 76.6 & 88.9 \\
News     & 75.2 & 74.2 & 74.0 & 71.3 & 84.9 & 94.0 & 70.4 & 71.1 & 69.1 & 68.6 & 82.6 & 92.7 \\
Prod.    & 90.6 & 78.5 & 89.1 & 75.9 & 94.2 & 95.8 & 84.7 & 74.2 & 83.2 & 71.6 & 91.3 & 93.4 \\
Network  & 66.8 & 73.9 & 65.9 & 71.2 & 80.1 & 93.5 & 62.7 & 70.5 & 61.5 & 67.9 & 78.6 & 91.2 \\
Tools    & 80.9 & 79.4 & 79.6 & 76.8 & 82.8 & 96.4 & 71.3 & 75.8 & 69.8 & 72.9 & 75.0 & 94.1 \\

\midrule
\textbf{Avg}
& 73.1 & 73.1
& 72.0 & 70.3
& 81.9 & 93.2
& 66.1 & 69.7
& 64.9 & 67.0
& 78.2 & 91.1 \\
\bottomrule
\end{tabular}}
\vspace{-0.4cm}
\end{table*}

\textit{Manual check.}
To ensure that automated metrics reflect actual migration correctness,
we manually inspected all migrated layouts produced in our evaluation.
Each case was examined against three criteria derived from our semantic model:
(i) preservation of view hierarchy and containment,
(ii) consistency of key attributes (e.g., text, size, and bindings), and
(iii) retention of essential layout relations such as alignment and weight distribution.
The inspection confirmed that layouts with high PSC scores consistently satisfied these semantic checks, whereas lower-PSC cases typically involved localized degradations such as flattening nested containers or approximating weighted layouts with heuristic spacers.
For example, in several generative transformations a \texttt{LinearLayout} with proportional weights (0.3/0.7) was visually reproduced but structurally rewritten as independent stacks, preserving appearance while altering layout semantics.
This check supports that PSC and related metrics align with human judgment of migration correctness.

\subsection{RQ3: How does \appname reduce manual development effort in UI transpilation?}
To evaluate the practical benefits of \appname for developers, we conducted a controlled user study comparing manual SwiftUI development against \appname-assisted migration. 
Due to the prohibitive cost of migrating all 31 applications manually, we randomly sampled 40 XML files drawn from 15 applications across eight domains, comprising 138 layouts and 368 view components (3,948 lines of UI code).

\begin{wraptable}{r}{0.5\textwidth}
\centering
\caption{
Time costs and proportion of each step in UI migration by \appname across dataset domains. 
}
\rowcolors{2}{gray!10}{white}
\begin{tabular}{lcc}
\toprule
\textbf{Group} & \textbf{LOC Changed} & \textbf{Time Cost} \\
\midrule
Control & 5,294 & 412 min \\
Experimental & 82 & 38 min \\
\bottomrule
\end{tabular}
\label{tab:development_effort}
\end{wraptable}

We recruited four professional mobile developers (3–5 years of experience, proficient in SwiftUI) and divided them into two groups:  
\textit{Experimental Group} (2 participants), who used \appname to generate SwiftUI code and only corrected migration errors; and  
\textit{Control Group} (2 participants), who implemented the same UIs manually in SwiftUI from scratch.  
Both groups were provided with the Android source code, UI screenshots, and pre-adapted iOS resources to ensure fairness.  

Each participant was required to produce SwiftUI code rendering an iOS interface visually consistent with the original Android design. 
The experimental group validated and refined \appname’s generated code, while the control group manually wrote all layout/view code and migrated resources by hand. 
We verified the final code in Xcode to ensure visual fidelity.

Table~\ref{tab:development_effort} reports development effort in terms of \emph{time cost} and \emph{lines of code (LOC) changed}, two widely used proxies for developer effort. 
On average, the experimental group completed each interface with only 82 LOC changes and 38 minutes of work, mainly involving localized refinements such as reconnecting project-specific resources, replacing unsupported widgets with native equivalents, and minor style adjustments rather than rewriting layout structures.
In contrast, the control group required 5,294 new lines of SwiftUI code and 412 minutes in total. 
This corresponds to over \textbf{90\% reduction in both code churn and development time} when using \appname.
The dramatic efficiency gains stem from three factors:  
(i) \appname automates resource parsing and adaptation, removing the need for manual asset migration;  
(ii) semantic transpilation generates structurally faithful SwiftUI code, minimizing manual layout reconstruction; and  
(iii) error handling localizes required fixes to a small set of unsupported components. 

Although minor manual adjustments remain necessary, they are significantly less costly than reimplementing entire UIs.
\appname reduces development time and effort by an order of magnitude compared with manual development, demonstrating its strong practicality for cross-platform UI migration in real-world scenarios.

\begin{figure}
\centering
\includegraphics[width=0.85\linewidth]{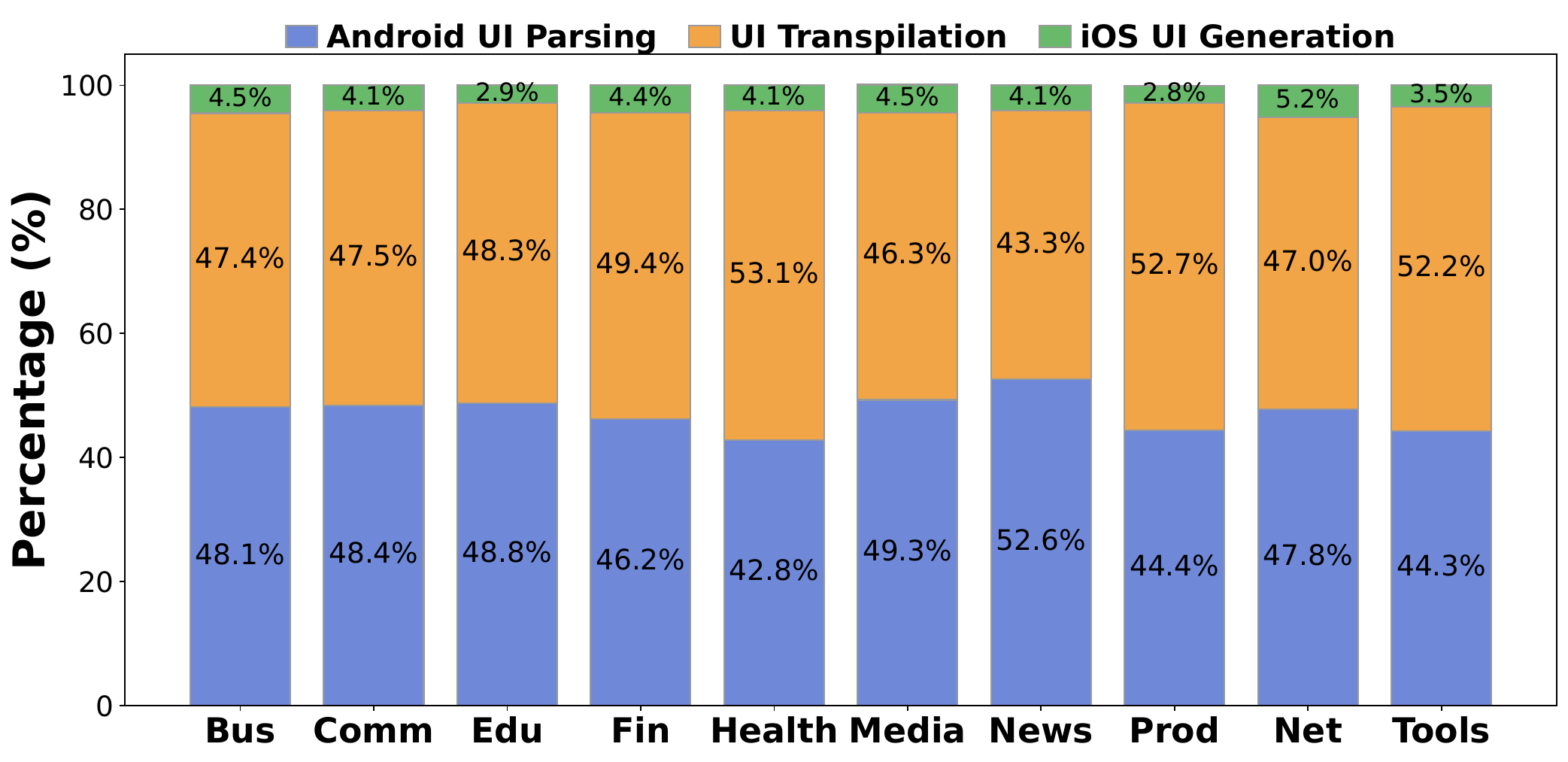}
\caption{Time costs and proportion of each step in UI migration by \appname on each dataset domain.}
\label{fig:time}
\end{figure}

\subsection{RQ4: What is the runtime efficiency of \appname in UI migration?}
To evaluate performance, we measure the end-to-end migration time of \appname across all 31 applications, and break down the cost into three stages: (i) UI parsing and resource adaptation, (ii) semantic transpilation, and (iii) target code generation. 
For comparison, we also record the time required by a GPT-5–based LLM workflow, including both API invocation and inference latency.

Figure~\ref{fig:time} reports the distribution of runtime costs across domains. 
On average, \appname migrates one complete application in \textbf{8.1 seconds}, showing that the pipeline is lightweight and efficient. 
Among the three stages, semantic transpilation accounts for the largest portion (\textasciitilde49\%), followed by parsing and resource adaptation (\textasciitilde47\%), while code generation contributes less than 4\%. 
The overall runtime scales linearly with the number of XML files and views, confirming the predictable behavior of the rule-based transformation process.

In contrast, the GPT-5–based workflow requires substantially more time per application, mainly due to inference latency and repeated API interactions.
This highlights the efficiency advantage of a compiled transpilation pipeline for application-scale migration.
Overall, \appname achieves efficient migration at the application level, with low runtime overhead and predictable scaling, making it suitable for integration into real-world development workflows.


%% file: discussion.tex
\section{Discussion}
\label{sec:discussion}

\subsection{LLMs for GUI Migration}
Our experiments with recent LLMs (ChatGPT-5.2 and DeepSeek-V3.2) show both the potential and limitations of generative approaches for cross-platform UI migration.
On the positive side, LLMs achieve moderate visual similarity, confirming their ability to generate plausible UI code given natural language prompts and structured inputs. 
However, we also observe several characteristic failure modes: (i) Structural drift. LLMs often restructure layouts in unintended ways, such as splitting elements into submodules or introducing undeclared variables, which reduces fidelity and can trigger compilation errors. (ii) Stylistic hallucination. LLMs may autonomously add modifiers (e.g., rounded corners, shadows) not present in the source, thereby altering the intended appearance. 
(iii) Syntax instability. Generated code may reference unsupported properties or undefined components, necessitating manual correction. 

In contrast, \appname takes a deterministic, semantics-preserving approach. 
By explicitly modeling layout semantics and applying platform-specific transformations, it avoids the probabilistic drift of LLMs, yielding higher structural fidelity and more stable code generation.

\subsection{Exploratory Comparison with Generative Migration Approaches}
\label{discussion_exploratory}
To understand the applicability of LLM-based agents and multimodal screenshot-to-code models to UI migration, we conducted a small-scale exploratory study to evaluate whether these generative approaches preserve the structural semantics required for migration.
We compared \appname with two representative paradigms:
(1) an agent-based workflow using Codex-5.2~\cite{codex}, where XML layouts and referenced resources were provided to generate Compose and SwiftUI code; and
(2) a vision-driven workflow using a multimodal ChatGPT-5.2~\cite{chatgpt}, where original Android UI screenshots were given as input to generate corresponding implementations in the target declarative framework.
We sampled five projects and ten representative layouts per project (100 layouts total) and evaluated outputs using the same metrics as RQ2 (SSIM, PSC, and compilation validation).

\begin{wraptable}{r}{0.5\textwidth}
\centering
\caption{Exploratory comparison with generative approaches on a sampled subset (100 layouts).}
\label{tab:llm_compare}
\begin{tabular}{lccc}
\toprule
\textbf{Approach} & \textbf{SSIM} & \textbf{PSC}  & \textbf{Err.}  \\
\midrule
Codex-5.2   & 0.81 & 0.72 & 3 \\
ChatGPT-5.2 (vision)  & 0.84 & 0.65 & 6 \\
\appname                 & 0.83 & 0.89 & 5 \\
\bottomrule
\end{tabular}
\end{wraptable}

From Table~\ref{tab:llm_compare}, we can see that both generative approaches produce visually plausible interfaces, achieving SSIM comparable to \appname.
However, their PSC scores are lower, indicating weaker preservation of component structure, attribute bindings, and layout relationships.
This behavior likely stems from the fact that generative models synthesize UI code without enforcing a shared cross-file semantic representation, whereas \appname performs semantic transpilation driven by an explicit intermediate representation.
These observations indicate that generative synthesis and semantic transpilation serve complementary roles:
LLM-based approaches enable rapid UI reconstruction, while transpilation provides stronger guarantees of structural consistency required for reliable large-scale migration.

\subsection{Extensibility and Maintenance Cost}
A common concern for migration approaches is whether supporting new UI constructs incurs substantial manual effort as frameworks evolve.
To assess this in practice, we implemented support for additional UI features beyond the original prototype, including modern layout patterns, interaction widgets, and less common attribute semantics.
As summarized in Table~\ref{tab:extension_cost}, these extensions required only localized changes at the adapter and rendering layers, typically on the order of tens of lines of code per category.
No modification to the intermediate representation, resource normalization, or core transpilation pipeline was required.

\begin{wraptable}{r}{0.66\textwidth}
\centering
\footnotesize
\setlength{\tabcolsep}{5pt}
\renewcommand{\arraystretch}{1.05}
\caption{Effort for extending \appname to additional UI constructs.}
\label{tab:extension_cost}
\begin{tabular}{lcl}
\toprule
\textbf{Category} & \textbf{Added LOC} & \textbf{Representative Extensions} \\
\midrule
Layout      & 34 LOC & MotionLayout, CollapsingToolbarLayout \\
View        & 41 LOC & RecyclerView, NavigationRailView, SearchView \\
Attribute  & 27 LOC & foregroundGravity, measureWithLargestChild \\
\bottomrule
\end{tabular}
\end{wraptable}

These observations indicate that extending \appname primarily involves refining how existing semantic roles are instantiated for new components, rather than introducing new translation mechanisms.
When unsupported widgets are encountered, \appname preserves structural layout information through a fallback representation, allowing developers to perform lightweight post-adjustments without disrupting the migrated hierarchy.

\subsection{Threats to Validity}
\textbf{Internal validity.} 
Our evaluation focuses primarily on UI components provided by the official Android framework. 
For uncommon or advanced components (e.g., \texttt{TextureView} for GPU-intensive apps), \appname currently substitutes placeholders rather than providing exact transpilation. 
While this may slightly underestimate migration fidelity, the underlying transpilation architecture is modular and allows developers to extend coverage by implementing custom rules for such components.
\textbf{External validity.} 
The generalizability of our results may be affected by applications that rely heavily on third-party or customized UI components. 
When no equivalent construct exists in the target platform (e.g., \texttt{com.kyleduo.switchbutton.SwitchButton}, which supports multi-state animations not directly available in SwiftUI), \appname simplifies the component to the closest standard alternative and logs the case in a migration report. 
This ensures functional completeness but may reduce visual fidelity. 
Although this limitation is inherent to cross-platform UI differences, our reporting mechanism enables developers to identify and refine such cases manually.
We mitigate these threats by (i) selecting a diverse dataset covering multiple domains, (ii) explicitly reporting unsupported components and their locations, and (iii) designing the transpiler to be easily extensible to new UI elements. 
Together, these measures reduce the risk of overstating the effectiveness of \appname.


%% file: related_work.tex
\section{Related Work}
\label{sec:relatedwork}
\noindent \textbf{Cross-platform Mobile Development Frameworks}.
Cross-platform development frameworks have become increasingly popular, facilitating the creation of mobile applications that can run on multiple platforms, e.g., Android and iOS.~\cite{biorn2020empirical,el2017taxonomy,heitkotter2013cross,el2016enhanced,ettifouri2017toward,biorn2018bridging,chadha2017facilitating,nunkesser2018beyond,Rieger2019TowardsTD,BirnHansen2018ASA,Karami2023OnTI,Lachgar2022DecisionFF}.
Wafaa S. et al.~\cite{el2016enhanced} developed a novel code transformation technique leveraging XSLT and regular expressions. 
This approach aimed to streamline the cross-platform mobile development process, facilitating the conversion of applications from Windows Phone 8 to Android. 
Henning et al.~\cite{heitkotter2013cross} introduced MD2, a model-driven approach for cross-platform application development. 
This approach allowed developers to define applications using a concise domain-specific language (DSL). 
From the abstract model, the system could automatically generate applications for both Android and iOS.
However, while cross-platform technologies offer certain advantages, they still fall short compared to native development, particularly in terms of performance, support for platform-specific features, and application size.


\noindent \textbf{Screenshot-Based UI Generation}.
Generating UI code from screenshots or UI design diagrams is a popular research topic~\cite{chen2018ui,xiao2024prototype2code,beltramelli2018pix2code,natarajan2018p2a,si2024design2code,nguyen2015reverse,liu2018improving,chen2019storydroid,chen2019automated}. 
Typically, these approaches use computer vision to identify the UI structure within images and leverage deep learning techniques to classify UI elements, thereby constructing the corresponding UI code.
Beltramelli et al.~\cite{beltramelli2018pix2code} introduced pix2code, an approach based on convolutional and recurrent neural networks. 
Their approach took a single screenshot as input and generated the GUI using a domain-specific language (DSL) to describe UI elements and reduce the search space. 
While pix2code generated iOS UIs based on Storyboard from screenshots, it could not produce declarative code. 
Natarajan et al.~\cite{natarajan2018p2a} presented the P2A tool, which combined optical character recognition (OCR) technology and heuristic rules to identify UI elements in Android screenshots. 
Screenshot-based UI generation technologies were constrained by the accuracy of computer vision in recognizing images, making it challenging to handle complex graphical structures.
Additionally, deep learning-based approaches require substantial training data, which is difficult to obtain for the iOS platform due to the lack of open-source data.
Furthermore, no current technology can generate declarative UI structures.
\appname addresses this gap by using migration techniques to facilitate cross-platform reuse of UI.

\noindent \textbf{UI Testing Migration}. 
There is extensive research on the migration of mobile testing cases~\cite{talebipour2021ui,qin2019testmig,behrang2020apptestmigrator,ji2023vision,lin2019test,lin2022gui,behrang2019test,behrang2018automated}. 
They aim to enable the reuse of test cases across applications with the same or similar functionalities, which can include cross-platform scenarios. 
By leveraging existing test cases for new or related applications, these approaches seek to optimize testing processes and improve efficiency, as well as reduce redundancy and save efforts.
Mariani et al.~\cite{mariani2018augusto} framed the problem of test reuse as a search challenge, employing evolutionary testing to facilitate test migration between different Android applications. 
TestMig~\cite{qin2019testmig} is a migration tool designed to transfer tests from iOS to Android by leveraging UI events. 
This tool requires access to the source code of both the source and target applications. 
Compared to the migration of UI test cases, our \appname focuses on the migration of the UI code itself, enabling cross-platform reuse of UI components.

%% file: conclusion.tex
\section{Conclusion and Future Work}
\label{sec:conclusion}

We presented \appname, a semantics-preserving approach for mobile UI migration that enables the reuse of Android XML-based interfaces on declarative frameworks such as SwiftUI and Jetpack Compose. 
Our evaluation across 31 applications from ten domains shows that \appname achieves high migration completeness and visual fidelity, while reducing manual development effort by more than an order of magnitude compared with manual coding, and outperforming LLM-based baselines.
In future work, we plan to broaden the scope of migration by supporting bidirectional transformations and additional platforms (e.g., Flutter, React Native). 
We further plan to extend \appname to support richer UI semantics (e.g., animations, accessibility tags, interaction logic, and dynamically constructed UI elements in Java/Kotlin through snapshot or runtime analysis).


\section{Data Availability}
Our tool is available at~\cite{uimigrator}.